\documentclass[reprint,notitlepage,aps,pra,superscriptaddress,amsmath,amssymb]{revtex4-1}

\usepackage{graphicx}
\usepackage{dcolumn}
\usepackage{bm}
\usepackage{amsmath}
\usepackage{color}
\usepackage{textgreek}
\usepackage[utf8x]{inputenc}
\usepackage{enumerate}
\usepackage{bibunits}
\usepackage{xspace}
\usepackage{pifont}
\allowdisplaybreaks[2]

\newcommand{\ket}[1]{|#1\rangle}				
\newcommand{\bra}[1]{\langle #1|}				
\newcommand{\ew}[1]{\langle #1 \rangle}				
\newcommand{\ketbra}[1]{| #1 \rangle \langle #1 |}		

\newcommand{\hide}[1]{}

\renewcommand{\textalpha}{$\alpha$}
\renewcommand{\textmu}{$\mu$}

\newcommand{\mone}{\ding{192}\xspace}
\newcommand{\mtwo}{\ding{193}\xspace}
\newcommand{\mthree}{\ding{194}\xspace}

\begin{document}

\title{
Coherent x-ray-optical control of nuclear excitons with zeptosecond phase-stability
}

\author{K. P. Heeg}
\affiliation{Max-Planck-Institut f\"{u}r Kernphysik, Saupfercheckweg 1, 69117 Heidelberg, Germany}

\author{A. Kaldun}
\affiliation{Max-Planck-Institut f\"{u}r Kernphysik, Saupfercheckweg 1, 69117 Heidelberg, Germany}

\author{C. Strohm}
\affiliation{Deutsches Elektronen-Synchrotron DESY, Notkestra{\ss}e 85, 22607 Hamburg, Germany}

\author{C. Ott}
\author{R. Subramanian} 
\author{D. Lentrodt}
\affiliation{Max-Planck-Institut f\"{u}r Kernphysik, Saupfercheckweg 1, 69117 Heidelberg, Germany}

\author{J. Haber}
\author{H.-C. Wille}
\affiliation{Deutsches Elektronen-Synchrotron DESY, Notkestra{\ss}e 85, 22607 Hamburg, Germany}

\author{S. Goerttler}
\affiliation{Max-Planck-Institut f\"{u}r Kernphysik, Saupfercheckweg 1, 69117 Heidelberg, Germany}

\author{Rudolf R\"uffer}
\affiliation{ESRF-The European Synchrotron, CS40220, 38043 Grenoble Cedex 9, France}

\author{C. H. Keitel}
\affiliation{Max-Planck-Institut f\"{u}r Kernphysik, Saupfercheckweg 1, 69117 Heidelberg, Germany}

\author{R. R\"ohlsberger}
\affiliation{Deutsches Elektronen-Synchrotron DESY, Notkestra{\ss}e 85, 22607 Hamburg, Germany}

\author{T. Pfeifer}
\author{J. Evers}
\email{joerg.evers@mpi-hd.mpg.de}
\affiliation{Max-Planck-Institut f\"{u}r Kernphysik, Saupfercheckweg 1, 69117 Heidelberg, Germany}

\date{\today}

\maketitle
\allowdisplaybreaks[2]
\textbf{
Coherent control of quantum dynamics is key to a multitude of fundamental studies and applications alike~\cite{Shapiro}. In the visible or longer-wavelength domains, near-resonant light fields have become the primary tool to control electron dynamics~\cite{Mukamel1995}. Recently, coherent control in the extreme-ultraviolet range was demonstrated~\cite{Prince2016}, with timing stability of the applied light fields in the few-attosecond range. At hard x-ray energies, M\"ossbauer nuclei feature narrow nuclear resonances, and spectroscopy of these resonances is a widespread tool to study magnetic, structural and dynamical properties of matter~\cite{moessbauer_story,Roehlsberger2005}. It has been shown that the power and scope of M\"ossbauer spectroscopy can be significantly advanced using various control techniques~\cite{Shvydko1996,Helistoe1991,Schindelmann2002,Vagizov2014,Heeg2017,Bocklage2017,Roehlsberger2010,Roehlsberger2012,Heeg2015b,Vagizov2013,Sakshath2017}. 
However, the coherent control of atomic nuclei using near-resonant x-ray fields remains an open challenge, also because of the extreme stability requirements on the x-ray light in the few-zeptosecond range. 
Here, we demonstrate such control, and use the relative phase of two x-ray pulses to switch the nuclear dynamics between stimulated emission and enhanced coherent excitation. 
For this, we suggest and implement a method to shape single pulses delivered by modern x-ray facilities into tunable double-pulses, with the desired stability on the few-zeptosecond level. 
Our results unlock coherent optical control for nuclei, and pave the way for  nuclear Ramsey spectroscopy~\cite{Ramsey1950} and spin-echo-like techniques,  which not only provide key concepts for advancing nuclear quantum optics~\cite{Adams2013}, but also essential ingredients for possible x-ray clocks and frequency standards~\cite{Riehle2006}. As a long-term perspective, we envision time-resolved studies of nuclear out-of-equilibrium dynamics, which is a long-standing open challenge in M\"ossbauer science~\cite{Shenoy2008}.
}

Coherent control refers to the control of quantum dynamics by light, based on coherence and interference phenomena~\cite{Shapiro,Mukamel1995}. In this process, central requirements are the capability to shape light pulses, and to stabilize the light's phase to a fraction of the oscillation period of its electric field. For M\"ossbauer nuclei, the relevant $E\sim 10$~keV photon energy range corresponds to an oscillation period on the hundred-zeptosecond time scale ($h / E \sim 400$~zs), such that x-ray-optical coherent control  requires stabilization to the few-zeptosecond scale, which has not been reported yet.

\begin{figure}[b]
 \centering
 \includegraphics[width=\columnwidth]{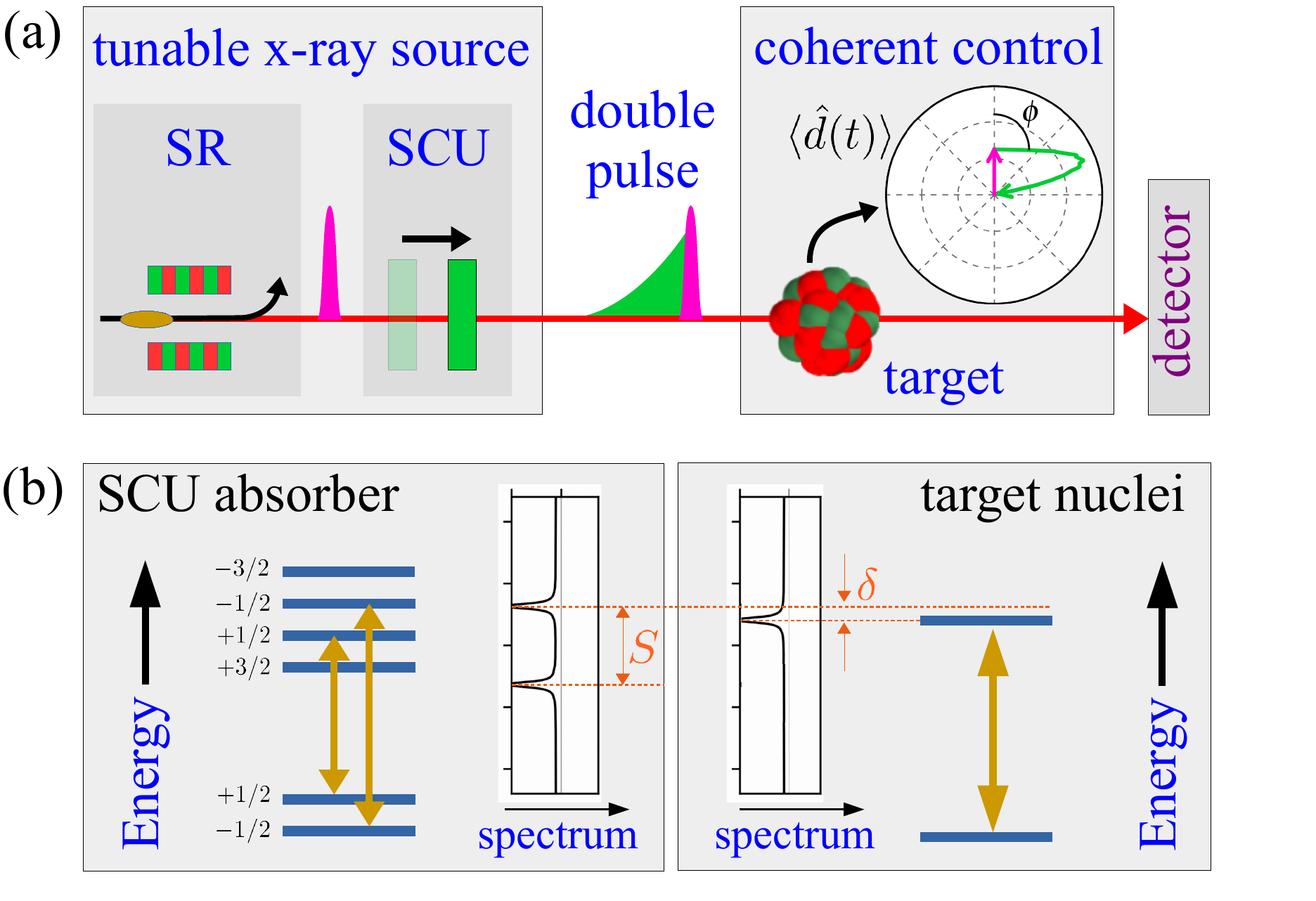}
 \caption{{\bf Schematic setup and samples.} (a) A short synchrotron (SR) x-ray pulse is shaped into a double pulse using a resonant absorber acting as a delay stage, which we denote as split-and-control unit (SCU). A fast motion of the SCU controls the relative phase $\phi$, detuning, and chirp of the two pulses, thus forming a tunable x-ray double-pulse source.
 The double-pulses are used to coherently control the dynamics of the target nuclei. An exemplary  dynamics is visualized via the nuclear dipole moment $\langle \hat{d}(t)\rangle$ on a polar plot, with a relative phase $\phi$ between the two pulses.
 (b) Energy level schemes and spectra of SCU absorber and target nuclei. For the coherent control, we tune the single resonance of the target nuclei to one of the two resonances of the SCU absorber ($\delta = 0$). 
}
 \label{fig:setup}
\end{figure}

\begin{figure*}[t]
 \centering
 \includegraphics[width=0.95\textwidth]{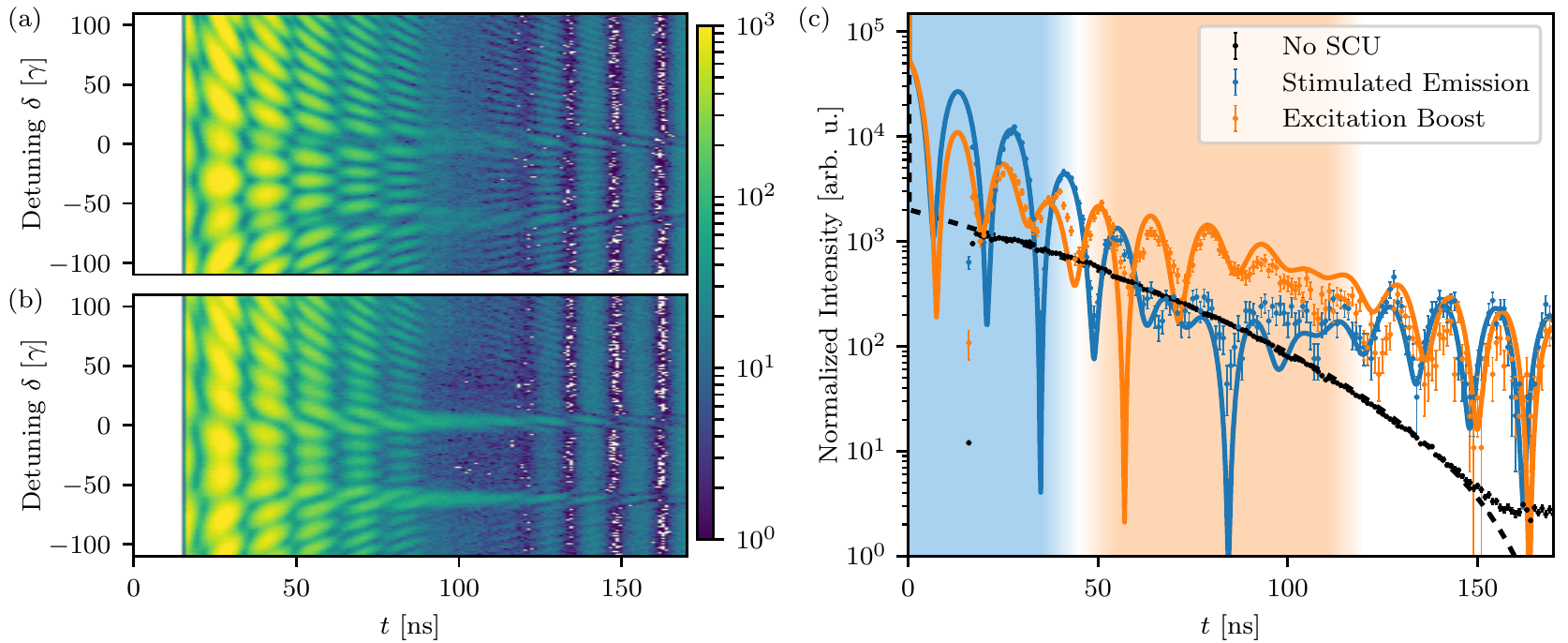}
 \caption{{\bf Experimental observation of the coherent control. }
 Time-and energy-resolved intensities recorded in forward direction for two different double-pulse sequences corresponding to stimulated emission (a) and enhanced coherent excitation (b) of the target nuclei, respectively. (c) Time-dependent intensity at relative detuning $\delta = 0$, normalized to equal measurement times. The experimental data (dotted) exhibits the characteristic crossover  (shaded areas) in the count rate between the two control cases at $\approx 45$~ns. Corresponding theory curves are shown as lines.}
 \label{fig:exp}
\end{figure*}

In previous works, it was demonstrated that incoherent light or conversion electrons enable one to study the excitation dynamics of nuclei, e.g., to reveal polariton propagation~\cite{PhysRevA.76.043811} or radiation trapping~\cite{PhysRevB.56.R8455}. These works are concerned with the nuclear population dynamics, but did not consider the control thereof or the phases characterizing the nuclear quantum state. Fast control of nuclear dynamics was  demonstrated, e.g.,  using sudden rotations of a static external magnetic field~\cite{Shvydko1996}, which allows for select control operations in sample materials with fast magnetic switching capabilities.
Another line of research involves rapid mechanical motions of one or more resonant absorbers as a means of control. This approach has been used to study polariton dynamics~\cite{PhysRevA.71.023804}, and in particular also to favorably shape x-ray pulses in the temporal~\cite{Helistoe1991,Schindelmann2002,PhysRevA.87.013807,Vagizov2014} or spectral domain~\cite{Heeg2017}. 
The latter works established the possibility to exploit this shaped x-ray light as a tool. While such pulse shaping techniques are reminiscent of their lower wavelength counterparts, which are essential for the realization of coherent control, the phase stability of the x-ray pulse shaping has not yet been evaluated and coherent control of nuclear quantum dynamics is yet to be demonstrated.

Here, we demonstrate the coherent control of the dynamics of M\"ossbauer nuclei using x-ray light. For this, we shape tunable double-pulse sequences out of given incident x-ray pulses using the mechanical motion of a resonant absorber (see Fig.~\ref{fig:setup}). In the main part of the experiment, we use the first (excitation) pulse of such sequences to induce a nuclear exciton in the target, i.e.~a single excitation coherently distributed over a large ensemble of nuclei. Controlling the relative phase of the second (control) pulse then enables us to switch the subsequent target dynamics between enhanced coherent excitation and stimulated emission of the nuclear exciton. 
Using an event-based time- and energy-resolved detection scheme which provides access to full holographic information of the outgoing light, we experimentally access the time-dependent magnitude and phase of the target's dipole moment, and demonstrate the  few-zeptosecond stability of our control scheme.

The double pulses are generated using a split-and-control unit (SCU, see Fig.~\ref{fig:setup}), which delays part of the incident x-ray pulse using a resonant absorber. The non-delayed fraction forms the leading excitation pulse 
$E_\textrm{exc}(t)$. The second control pulse  $E_\textrm{control}(t)$ consists of the delayed part. While the overall phase of the double pulses inherits the random fluctuations of the incident x-rays, the relative phase between the two pulses is stable. The double pulse can be tuned using mechanical motions $x(t)$ of the SCU absorber immediately after the x-ray excitation, imposing an additional translational phase $\exp{[ikx(t)]}$ onto the control pulse, where $k$ is the x-ray wavenumber. Sudden displacements, linear motion and non-linear motions of the SCU translate into phase shifts, detunings, and chirps of the control pulse relative to the excitation pulse, respectively. Since the control pulse is spectrally narrow due to the slow temporal decay of the SCU's resonant absorber, one can selectively choose the nuclear transitions to be addressed and controlled in the target. Overall, synchrotron and SCU  together thus form a tunable source for x-ray double pulses.

We experimentally realized the coherent control of nuclear dynamics via tunable x-ray double-pulses at the Nuclear Resonance Beamline ID18 at ESRF (Grenoble)~\cite{Rueffer1996}, see Fig.~\ref{fig:setup}. The nuclear target was formed by a stainless-steel foil with thickness 1~{\textmu}m, enriched in the M\"ossbauer isotope \textsuperscript{57}Fe to 95\%, which features a nuclear transition at energy 14.4~keV with a width of $\hbar\gamma = 4.7$~neV and a lifetime of $1/\gamma = 141$~ns. This energy translates into an oscillation period of the carrier frequency of $T_0 = 287$~zs, such that a phase change by $\pi$ corresponds to a temporal shift by $T_0 / 2$. 
As the delay stage in the SCU, we used an \textalpha-iron foil with thickness 2~{\textmu}m, also enriched in \textsuperscript{57}Fe. A weak external magnet was used to align its internal hyperfine field, such that only the two $\Delta m = 0$ transitions with energy splitting of $S \approx 63\gamma$ were driven, see Fig.~\ref{fig:setup}(b). Due to these two transitions, the SCU generates a bichromatic control pulse. In addition to the SCU movement, we used Doppler shifts to scan the relative detuning $\delta$ of the resonance frequencies of the target nuclei and the SCU absorber.
The characterizations of the samples and the experimentally realized double-pulse sequences and SCU motions are described in Appendix~\ref{app-samples}.

\begin{figure}[t]
 \centering
 \includegraphics[width=\columnwidth]{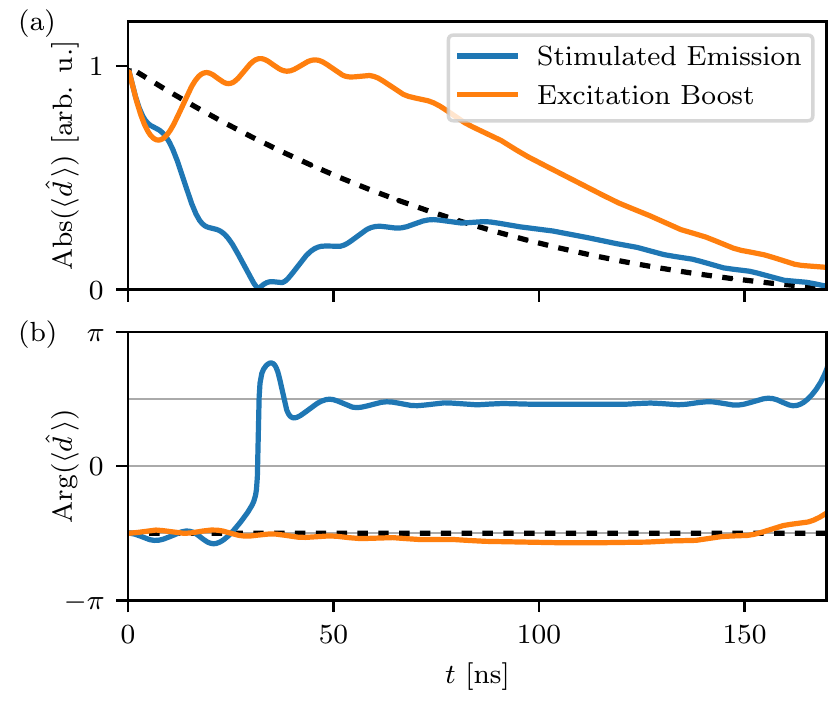}
 \caption{{\bf Time-dependent dipole moment of the target nuclei. }
 (a), (b) show the modulus and the phase of the nuclear dipole moment reconstructed from the experimental data. 
 The accelerated decay in the case of stimulated emission (blue) and the enhanced coherent excitation (orange) are clearly visible. A theoretical reference calculation without SCU is shown as black dashed line.}
 \label{fig:dipole}
\end{figure}

To demonstrate the coherent control of the target nuclei, we compared two different double-pulse sequences. In both cases, the short excitation pulse drives the nuclear ensemble into an excitonic state at $t = 0$~\cite{Hannon1999}. 
In the first sequence, the phases of the control and the excitation pulses coincide, such that a further coherent excitation of the nuclei due to the control pulse is expected ({\it enhanced coherent excitation}). In the second sequence, the control- and excitation pulses have opposite phases, and the control pulse is expected to drive the exciton created by the excitation pulse back to the ground state, corresponding to a {\it stimulated emission} of the excitonic state. 

The fluorescence light emitted by the target nuclei provides a direct experimental signature of the induced dynamics, as its amplitude is proportional to the nuclear dipole response $\ew{\hat{d}(t)}$. In forward direction, the fluorescence interferes with the driving double-pulse, resulting in a total intensity 
$I_\textrm{total}(t, \delta) =|  E_\textrm{exc}(t, \delta) + E_\textrm{control}(t, \delta) + \alpha \ew{\hat{d}(t, \delta)}  |^2$,
where $\alpha$ is a constant (see  Appendix~\ref{app-qom}). Recording this intensity as function of time and relative detuning $\delta$, allows us to exploit the interference to experimentally access the  complex nuclear dipole moment as an observable.

The recorded time- and energy-resolved intensity spectra for the two double-pulse sequences are shown in Fig.~\ref{fig:exp}(a,b). As a first result, we find that the two pulse sequences lead to substantially different spectra, which is most visible at the two  SCU absorber resonances around $\delta = 0\gamma, \, 63\gamma$.
A model-independent fit to the two-dimensional spectra allows us to determine the precise motion of the SCU~\cite{Heeg2017}, and thereby  the time-dependent field amplitude of the generated double pulses (see Appendix~\ref{app-scu}), setting the stage for the coherent control of the target nuclei.

To realize the coherent control of the target nuclei, we tuned the target nuclei in resonance with one of the SCU absorber's spectral lines (relative detuning $\delta = 0$ in Fig.~\ref{fig:setup}(b)), and measured time-dependent intensities in forward direction for the two motions. Results are shown in Fig.~\ref{fig:exp}(c), together with corresponding theory curves (see Appendix~\ref{app-scu}). By comparing  the two intensities, a characteristic crossover in the dominating intensity as a function of time is observed, which allows for a qualitative analysis of the dipole dynamics~\cite{Reichegger2014}. Initially, the intensity in the case of stimulated emission dominates, since at early times, the stimulated light is emitted in forward direction in addition to the incident light (blue shaded area in Fig.~\ref{fig:exp}(c)). Subsequently, the intensity for the enhanced coherent excitation case becomes dominant (orange shaded area), because of the increased excitation of the nuclei. In Appendix~\ref{app-cross}, we show that this characteristic intensity crossover indeed can be linked analytically to the two cases of spontaneous emission and enhanced coherent excitation.

For a quantitative analysis of the nuclear dynamics, we extract the dipole moment of the target nuclei from the experimental data (see Appendix~\ref{app-qom}). The results in Fig.~\ref{fig:dipole} clearly show  the effect of stimulated emission and enhanced coherent excitation, and agree well with corresponding model calculations (see Fig.~\ref{fig:patterns}). Without control pulse, the dipole moment exponentially decays, preserving its phase (black-dashed). In the stimulated emission case (blue), the control pulse rapidly and non-exponentially drives the nuclear excitation back to the ground state characterized by $|\langle \hat{d}\rangle|=0$ within about 30ns. Afterwards, the residual control pulse continues this dynamics through the ground state and re-excites the nuclei with opposite phase, before they exponentially decay after the end of the double-pulse sequence. 
In the enhanced coherent excitation case (orange), the control pulse significantly excites the magnitude of the dipole moment beyond the reference case without control pulse. The dipole phase is approximately constant, demonstrating that the control- and excitation pulse phases indeed agree. Note that the excitation increase starts a few ns after the initial excitation, because of the finite duration of the SCU's movement of about $15$~ns.

The importance of our multidimensional detection scheme is also highlighted by the comparison of Figs.~\ref{fig:exp} and \ref{fig:dipole} (see also Appendix~\ref{app-multi}). It demonstrates that the time-dependent intensity does not directly reflect the desired dynamics of the target nuclei, because of the interference between the incident pulse and the forward-scattered light~\cite{PhysRevA.76.043811}. In particular, the measured intensity in Fig.~\ref{fig:exp}(c) exhibits rapid oscillations. These so-called quantum beats~\cite{Roehlsberger2005} appear because the detector cannot individually resolve the two spectral components of the control pulse generated by the SCU, see Fig.~\ref{fig:setup}(b). In contrast, the dipole dynamics in Fig.~\ref{fig:dipole} only shows small residual oscillations, because the spectral response of the target nuclei is so narrow that they are   selectively driven by only one of the SCU's resonances, while the second SCU resonance is far-detuned.  We further note that because of this difference, we are not interested in optimizing the outgoing light in any respect, unlike previous works~\cite{Helistoe1991,Schindelmann2002,PhysRevA.87.013807,Vagizov2014,Heeg2017}. Rather, in our experiment, it acts as an experimental signature to observe the nuclear dynamics.

\begin{figure}[t]
 \centering
 \includegraphics[scale=1]{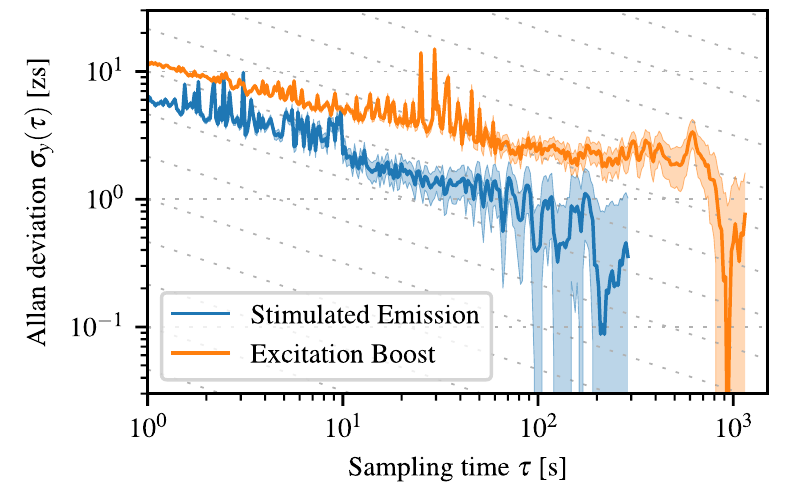}
 \caption{{\bf Stability of the double-pulse sequence.} The Allan deviation $\sigma_y(\tau)$ for both SCU operation modes is in the few-zs domain. The shaded areas show the error ranges, diagonal grid lines indicate $1/\sqrt{\tau}$ scaling.}
 \label{fig:allan}
\end{figure}

Key characteristics of coherent control schemes are their stability and reproducibility, which can be characterized via the Allan deviation $\sigma_y(\tau)$~\cite{Allan1966}. We analyze the stability of our coherent control scheme via the stability of the  SCU motion $x_0(t)$, to which we can attribute any perturbations, since only relative motions between SCU and absorber affect our results (see Appendix~\ref{app-allan} and Fig.~\ref{fig:allan_scheme}). We split the total measurement time into $N$ non-overlapping intervals of duration $\tau$, and analyze each interval $i$ separately. Because of the short duration of each x-ray pulse sequence (176~ns), the dominating noise is a linear drift which perturbs the SCU motion to $x_0(t) + A_i t$, where $A_i$ randomly fluctuates between intervals (see Appendix~\ref{app-allan}). We translate this drift into an upper bound for a temporal deviation $y_i = A_i\,t_2 / c$, where $t_2 = 170$~ns is the maximum range of our data acquisition, and $c$ the speed of light. Then, $\sigma_y(\tau) = [2 (N-1)]^{-1/2}[\sum_{i=1}^{N-1} (y_{i+1} - y_i)^2]^{1/2}$ characterizes the relative root mean square instability of two measurements $\tau$ apart. 
Results are shown in Fig.~\ref{fig:allan} as a function of $\tau$. As expected, the Allan deviation initially reduces with growing $\tau$, since noise is averaged out more effectively due to the increased statistics, thereby increasing the stability between successive measurements. At even longer times $\tau$, systematic drifts which are not removed by the $\tau$-averaging are expected to increase the Allan deviation again, but this regime is not clearly reached within our total measurement time.  We find that the stability reaches the few-zeptosecond scale, both with and without motion of the SCU. This temporal stability exceeds the best reported value achieved with XUV optical interferometers by two orders of magnitude~\cite{Laban2012,Prince2016}. Fluctuations visible at intermediate $\tau$ are due to dead times of our detection system (see Appendix~\ref{app-detector}).
We note that this analysis crucially relies on the full holographic capabilities of our two-dimensional detection scheme, since the time-dependent intensity studied in previous experiments alone is incapable of detecting the relevant deviations (see Appendix~\ref{app-multi}). Further, an event-based detection is required for the a-posteriori  binning of the data into different time intervals $\tau$.

In addition to the phase control reported here, our SCU scheme may also induce detunings or frequency chirps between the two pulses. Furthermore, the control pulse could be temporally delayed by storing the x-ray pulse in the SCU for a variable time, e.g., by means of magnetic switching~\cite{Shvydko1996}. Such a split-control-delay-unit (SCDU) would additionally be able to set the polarization of the control pulse~\cite{PhysRevLett.103.017401}.
The control also generalizes to stronger excitation of the nuclear ensemble, e.g., involving x-ray free-electron laser sources~\cite{Chumakov2018,Heeg2016M}, directly opening up the avenue to explore nuclear dynamics using x-ray-pump -- x-ray-probe techniques. Similarly, our approach could promote emerging visible-pump -- x-ray-probe schemes~\cite{Vagizov2013,Sakshath2017}.
The paradigm shift from controlling x-ray light to controlling nuclear matter, together with the coherent control capabilities  demonstrated here, form an indispensable gateway to engineer complex quantum states and to explore time-dependent phenomena with nuclei, as in the longer-wavelength domain~\cite{Zewail2000,Ullrich2012,Mukamel2013}. We in particular envision the study of nuclear out-of-equilibrium dynamics, which is a long- standing open challenge in M\"ossbauer science~\cite{Shenoy2008}.

\section*{Acknowledgements}
We acknowledge a consolidator grant from the European Research Council (ERC) (X-MuSiC-616783).
This work is part of and supported by the DFG Collaborative Research Centre ``SFB 1225 (ISOQUANT).''

\appendix

\renewcommand{\thefigure}{S\arabic{figure}} 
\setcounter{figure}{0} 

\begin{figure}[t]
 \centering
 \includegraphics[width=0.95\columnwidth]{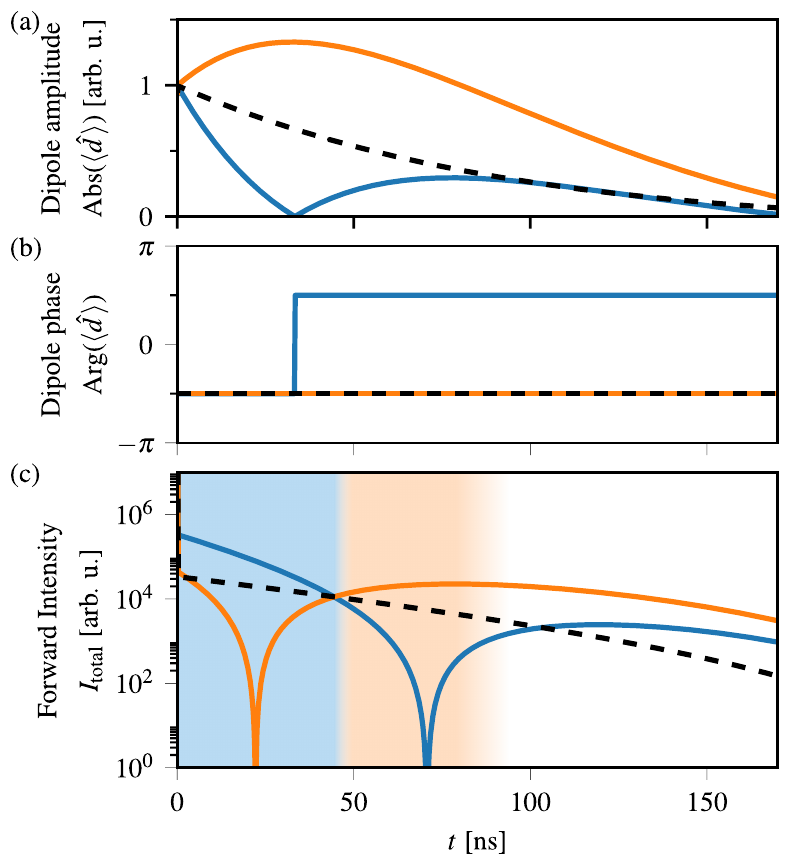}
 \caption{{\bf Theoretical predictions for the stimulated emission and enhanced coherent excitation of the nuclear exciton.}
 Nuclear dynamics under the action of different double pulses, visualized by the magnitude (a) and phase (b)  of the nuclear dipole moment. The excitation pulse induces a nuclear excitation at $t=0$, and the figures show the subsequent dynamics induced by the control pulse. Opposite phase between excitation and control pulse leads to stimulated emission, followed by subsequent coherent re-excitation (blue). Equal phase induces enhanced coherent excitation (orange). The black dashed line indicates the dipole response in the absence of the SCU.
 Panel (c) shows the total intensity emitted in forward direction.
  The shaded areas indicate a crossover in the dominating intensity.
 }
 \label{fig:patterns}
\end{figure}

\begin{figure}[t]
 \centering
 \includegraphics[width=0.95\columnwidth]{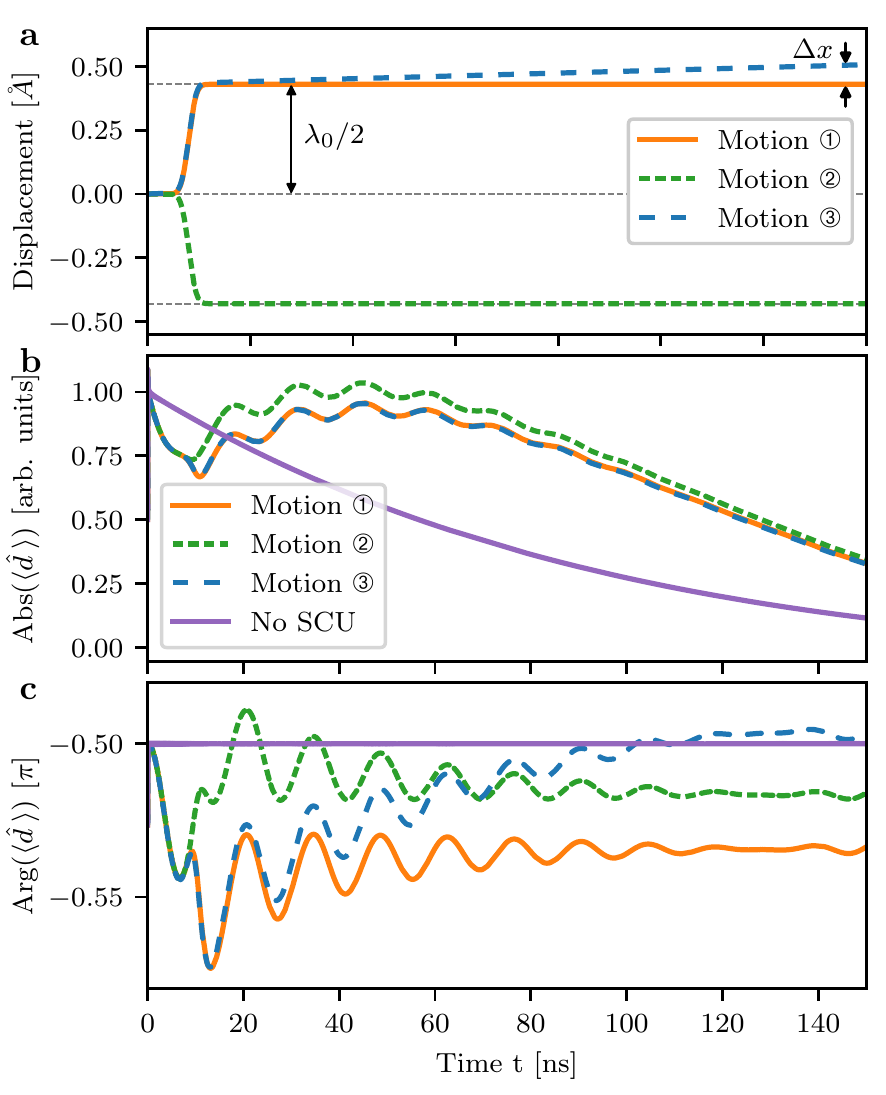}
 \caption{\textbf{Three absorber motions to illustrate the multidimensional detection.} (a) shows three SCU absorber motions used to illustrate the capabilities of the multidimensional detection. (b,c) show the magnitudes and phases of the dipole moments induced in the target nuclei due to the double pulses generated by the respective motions, or in the absence of an SCU.}
 \label{fig:multidim1}
\end{figure}

\begin{figure*}[t]
 \centering
 \includegraphics[width= 1.95\columnwidth]{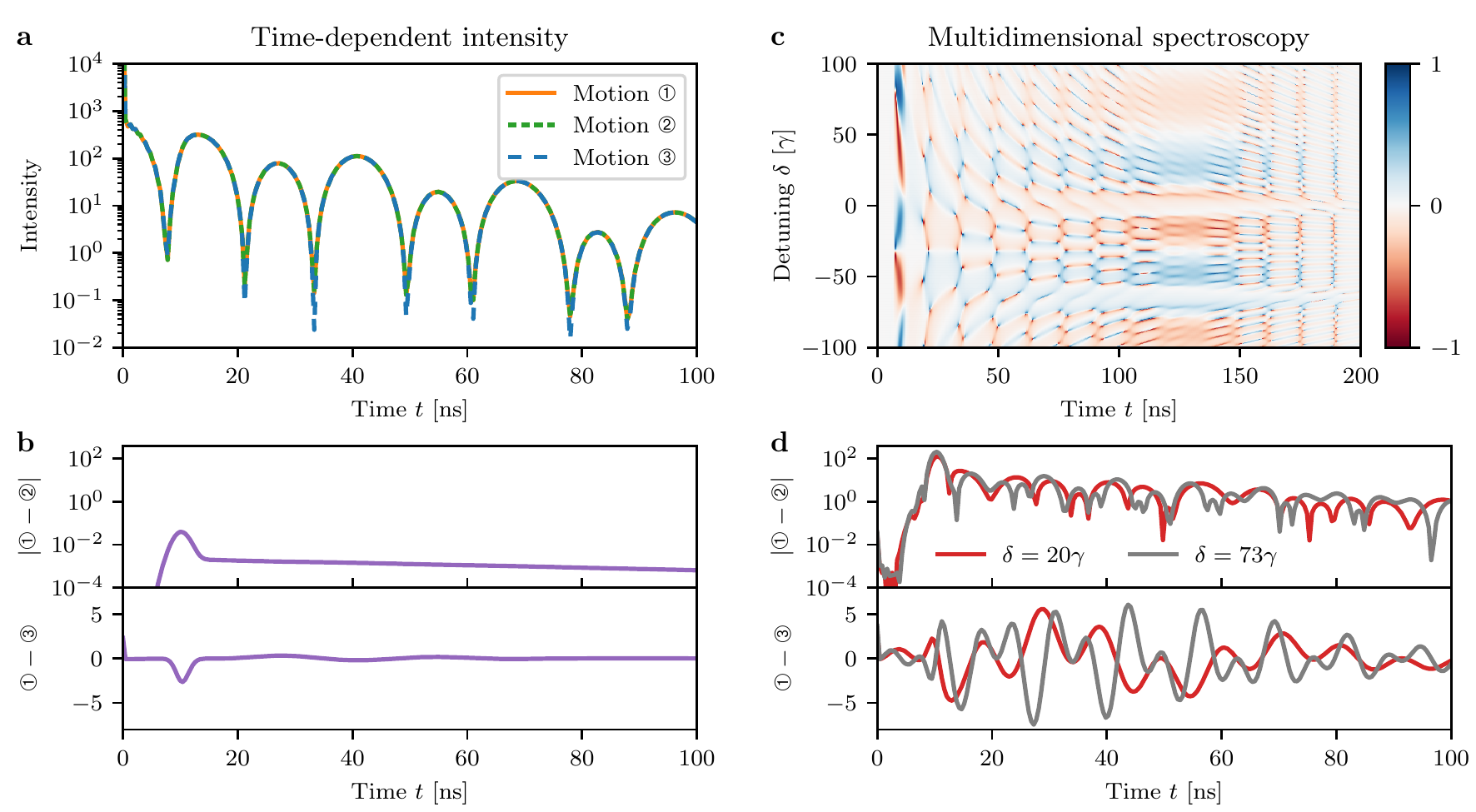}
 \caption{\textbf{Multidimensional detection and time-dependent intensities.} (a,b) show theoretical predictions for the resonant time-dependent intensity using the three motions in Fig.~\ref{fig:multidim1}. (a) demonstrates that these intensities for the three motions essentially coincide. (b) illustrates this further with the  differences between the time-dependent intensities for the motions \mone and \mtwo as well as  \mone and \mthree, respectively. 
 (c,d) show corresponding theoretical predictions for our multidimensional spectroscopy technique. (c)  shows the
 relative difference $(I_\text{\mtwo} - I_\text{\mone})/(I_\text{\mone} + I_\text{\mtwo})$ between the 2D spectra of motions \mone and \mtwo. The rich interference structures and high visibility show that the motions can clearly be distinguished. (d) shows intensity differences corresponding to the results in (b), but energy-resolved at  sections with different detunings $\delta$ through the 2D data, where $\delta$ is defined in Fig.~\ref{fig:setup} of the main text. Each detuning $\delta$ leads to characteristic strong interference patterns, further illustrating that the motions can easily be distinguished using the multidimensional detection.}
 \label{fig:multidim2}
\end{figure*}

\begin{figure}[t]
 \centering
 \includegraphics[width=0.95\columnwidth]{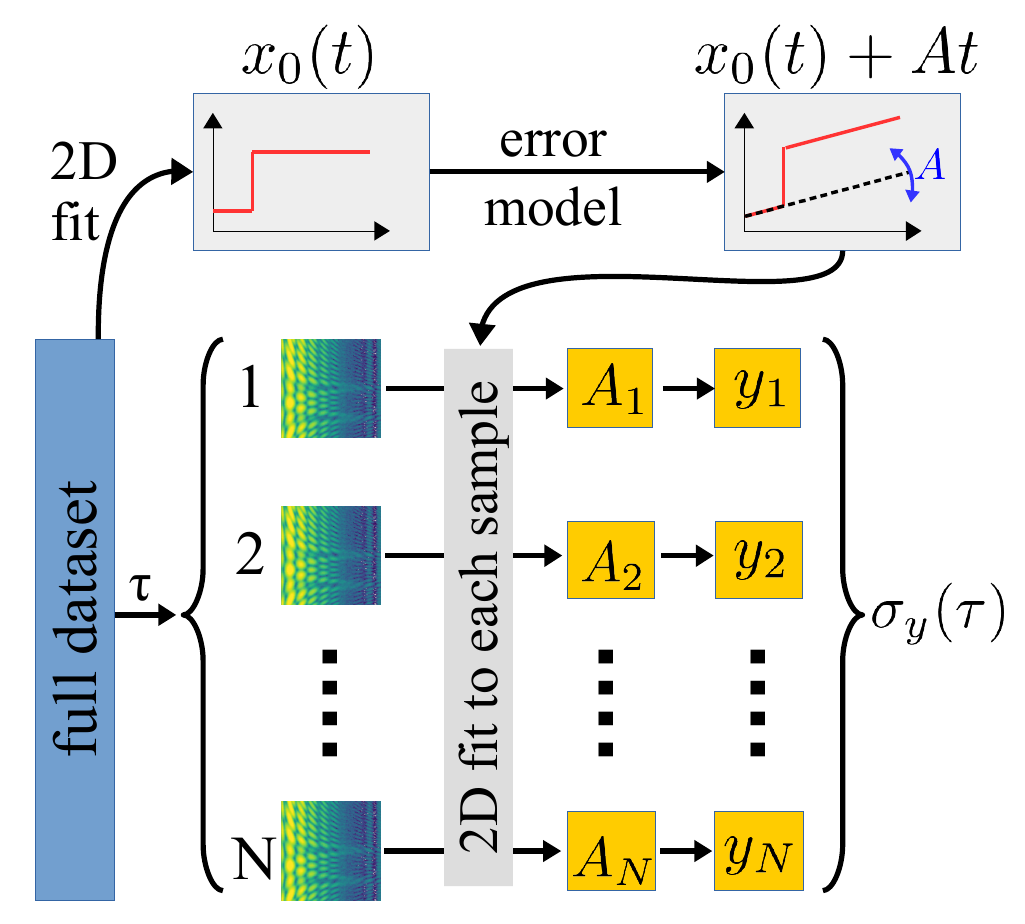}
 \caption{\textbf{Evaluation of the Allan deviation.} To evaluate the Allan deviation, the full dataset is split into $N$ samples of duration $\tau$. Each sample comprises a 2D time- and energy-resolved spectrum, to which we fit spectra obtained using the SCU motion $x_0(t)$ modified via an error model with a parameter $A$. This yields the best fit parameter $A_i$ for each sample $i$, which is proportional to the temporal error $y_i$ as explained in Appendix~\ref{app-allan}. From these $y_i$, the Allan deviation can be calculated using Eq.~(\ref{eq:allan}).}
 \label{fig:allan_scheme}
\end{figure}

\section{\label{app-nrs}Nuclear resonant scattering}
The forward transmission of an arbitrary x-ray pulse $E_\textrm{in}(t)$ in an extended resonant sample is~\cite{Kagan1979}
\begin{equation}
 E_\textrm{out}(t) = E_\textrm{in}(t) \ast T(t) \;, \label{eq:eoutnrs}
\end{equation}
where $T$ is a characteristic transmission function and ``$\ast$'' denotes a convolution.
Neglecting electronic absorption, one can write
\begin{equation}
 T(t) = \delta(t) + R(t) \;, \label{eq:transmission_delta_response}
\end{equation}
where $\delta(t)$ is the Dirac delta function and $R(t)$ denotes the response function of the nuclear target, i.e.~the scattered x-rays.
This response function $R(t)$ is directly related to the effective nuclear dipole $\ew{\hat{d}(t)}$, which forms the primary quantity of interest in this work.
For the relevant case of weak excitation and thin targets, we  analytically show this direct relation between the response function and the nuclear dipole below using a quantum optical two-level model.
Realistic transmission functions $T(t)$, which we use to model the experimental data, can be computed with software packages such as \textsc{conuss}~\cite{Sturhahn2000}.
The dispersive and absorptive properties of the electronic background are spectrally broad and are included as a constant factor.
For a material featuring a single-line resonance, $T(t)$ can be expressed analytically. Omitting the free phase evolution $\exp(i\omega_0 t)$, we have~\cite{Roehlsberger2005}
\begin{equation}
 T(t) = \delta(t) - \theta(t) \, \sqrt{\frac{b}{t}} \, e^{-\tfrac{\gamma}{2} t} \, J_1\left(2 \sqrt{b t} \right) \;. \label{eq:transmission_full}
\end{equation}
Here, $\theta(t)$ the Heaviside step function, $J_1$ the Bessel function of the first kind,
$b = \gamma \sigma_0 f nd / 4$ parameterises the thickness of the resonant target, $n$ is the volume density of the resonant nuclei, $d$ the target thickness, $\sigma_0$ the cross section, $f$ the Lamb-M\"ossbauer factor, and $\gamma$  the resonance width. For the data shown in Fig.~\ref{fig:patterns}, we used $b_\textrm{SCU} = 5 \gamma$ and $b_\textrm{Target} = 2.3\gamma$. These parameter choices also optimally mimic the experimentally realized setting.
For later comparison with the quantum optical model it is instructive to explicitly compute the scattered x-ray field in the limit of a thin target.
From Eq.~(\ref{eq:transmission_full}), one finds for small $b$ and short times $t$
\begin{equation}
 T(t) \approx \delta(t) - \theta(t) b e^{-\tfrac{\gamma + b}{2} t} \;, \label{eq:tthin}
\end{equation}
which indicates the well-known initial superradiant decay with rate $\gamma+b$~\cite{vanBuerck1992}.

\vskip 0.5cm
\noindent
{\bf Split-and-control unit (SCU) operation.}
Excited by a short $\delta(t)$-like x-ray pulse and in the stationary case, the field behind the SCU given in Eq.~(\ref{eq:eoutnrs})
reduces to Eq.~(\ref{eq:transmission_delta_response}).
In order to tune the relative phase between the $\delta(t)$ component and the scattered part $R(t)$, a motion $x(t)$ is applied to the SCU. This results (see also~\cite{Heeg2017}) in the combined field
\begin{eqnarray}
 E_\textrm{SCU}(t) &=& E_\textrm{exc}(t) + E_\textrm{control}(t) \nonumber \\
 &&= \delta(t) + e^{i\phi(t)} R(t)  \;, \label{eq:ein_phase1}\\
 \phi(t) &=& k \left[ x(t) - x(0) \right] \label{eq:ein_phase2} \;,
\end{eqnarray}
where $k = \omega_0 / c$ is the wavenumber.
In our experiment, we use this double pulse to drive a nuclear target. Again, the downstream x-ray intensity can be computed using Eq.~(\ref{eq:eoutnrs}),
where $E_\textrm{SCU}(t)$ now takes the role of the input field $E_\textrm{in}(t)$
and $T(t)$  corresponds to the transmission function of the actual target.

\section{\label{app-qom}Quantum optical two-level model}
In order to model the internal dynamics of the target by first principles, we use an approach based on a two-level-system (TLS) description for the resonant target. In the thin-sample limit and at weak excitation, it is known to agree with the nuclear resonant scattering approach from above. We exploit this equivalence to establish an expression for the target dipole moment.

The TLS is formed by one collective ground state $\ket{g}$ and one collective excited state $\ket{e}$. The driving with an x-ray field $E_\textrm{in}(t)$ is described by the Hamiltonian
\begin{equation}
 H = \frac{\Omega(t)}{2}\ket{e}\bra{g} + \frac{\Omega^*(t)}{2}\ket{g}\bra{e} \;,
\end{equation}
where $\Omega(t) = d \cdot E_\textrm{in}(t) / \hbar$ with $d$ being the dipole moment. Additionally,
we include spontaneous decay with rate $\tilde{\gamma}$ in terms of a density matrix
\begin{equation}
\frac{d}{dt}\rho = -i [H, \rho] -\frac{\tilde{\gamma}}{2} \left(\ketbra{e}\rho + \rho \ketbra{e} - 2 \ket{g}\bra{e} \rho \ket{e}\bra{g} \right) \;.
\end{equation}
For weak excitation it is sufficient to consider the coherence $\ew{\sigma_{ge}} = \bra{e}\rho\ket{g}$ only. In the limit $\bra{g}\rho\ket{g} = 1$, $\bra{e}\rho\ket{e} = 0$, we have the equation of motion
\begin{equation}
 \frac{d}{dt} \ew{\sigma_{ge}(t)} = -i \frac{\Omega(t)}{2}   - \frac{\tilde{\gamma}}{2} \ew{\sigma_{ge}(t)} \;,
\end{equation}
which is solved by (boundary conditions $\ew{\sigma_{ge}} = \Omega = 0$ at $t = -\infty$)
\begin{eqnarray}
 \ew{\sigma_{ge}(t)}
 &=& -\frac{i}{2} \int_{-\infty}^t      \Omega(t^\prime)                    e^{-\tfrac{\tilde{\gamma}}{2} (t - t^\prime) } dt^\prime \nonumber \\
 &=& -\frac{i}{2} \int_{-\infty}^\infty \Omega(t^\prime) \theta(t-t^\prime) e^{-\tfrac{\tilde{\gamma}}{2} (t - t^\prime) } dt^\prime \nonumber \\
 &=& -\frac{i}{2} \Omega(t) \ast \left[ \theta(t) e^{-\tfrac{\tilde{\gamma}}{2} t} \right] \;. \label{eq:sigma_t}
\end{eqnarray}
The field behind the TLS is composed of the initial field and a scattered contribution~\cite{Agarwal1974}
\begin{equation}
 E_\textrm{out}^\textrm{TLS}(t) = E_\textrm{in}(t) + \alpha \ew{\hat{d}(t)} \;, \label{eq:eoutqo}
\end{equation}
where $\ew{\hat{d}} = d \ew{\sigma_{ge}}$ is the dipole response, and $\alpha$ is a constant, also taking into account the extended sample geometry~\cite{Heidmann1985}.
In particular, for
$\alpha = -2 i b \hbar / d^2$ and $\tilde{\gamma} = \gamma + b$ we have
\begin{equation}
 E_\textrm{out}^\textrm{TLS}(t)
 =E_\textrm{in}(t) \ast \left[ \delta(t) - \theta(t) b e^{-\tfrac{\gamma + b}{2} t} \right] \;,
\end{equation}
which is the same result found in nuclear resonant scattering theory for thin samples, see Eqs.~(\ref{eq:eoutnrs}) and~(\ref{eq:tthin}).
The analytical agreement demonstrates the validity of the TLS approach.
Comparing Eq.~(\ref{eq:eoutqo}) with Eqs.~(\ref{eq:eoutnrs}) and~(\ref{eq:transmission_delta_response}),
we find
\begin{equation}
 E_\textrm{in}(t) \ast R(t)= \alpha \ew{\hat{d}(t)}\;,
\end{equation}
which highlights the correspondence of the response function in the nuclear resonant scattering approach with the time-dependent nuclear dipole in the quantum optical model.

\begin{figure}[t]
 \centering
 \includegraphics[width=0.95\columnwidth]{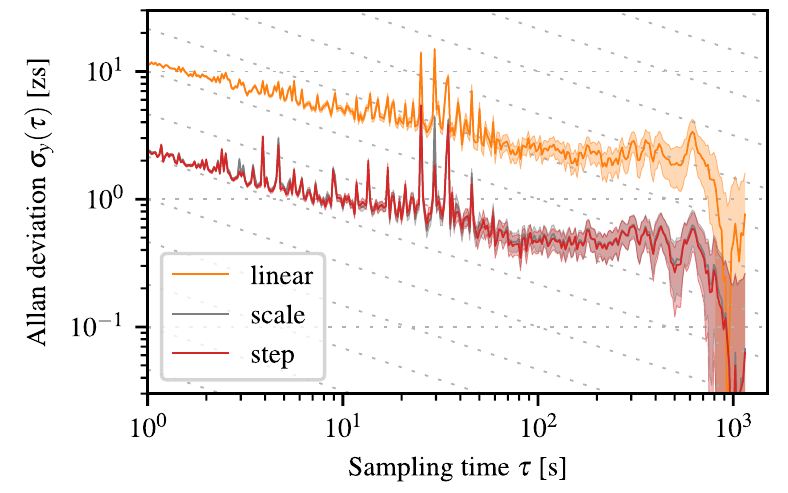}
 \caption{\textbf{Comparison of different noise models.} Allan deviation in case of enhanced coherent excitation obtained for the three employed noise models.
 The uncertainty according to the linear noise model exceeds that of other models by about a factor three. Shaded areas are error ranges, diagonal grid lines indicate $1/\sqrt{\tau}$ scaling.}
 \label{fig:allan_noisemodels}
\end{figure}

Finally, we note that our method is not restricted to thin targets. For example, one can  associate an effective dipole moment to the dynamics of a thicker sample, and to evaluate it from experimental data using our approach, or one can resolve the intra-target dynamics as function of the penetration depth. However, in the experiment we focus on the thin-target limit, because it allows us to directly observe the control of the nuclear dynamics from the experimental raw data using a clear intensity-crossover criterion discussed next.

\section{\label{app-cross}Intensity crossover}
When comparing the different SCU operations in the coherent control setting, differences are found in the temporal structure of the x-ray field behind the target (see Fig.~\ref{fig:exp}).
In particular, the most prominent qualitative feature for the cases considered here is a crossover of the dominating intensity after a certain time.
This behaviour is also predicted from our quantum optical model and equivalently, from the nuclear resonant scattering approach in the thin-sample limit.
In the case of stimulated emission (SE), no displacement of the SCU is required. The double pulse is given by (see Eq.~(\ref{eq:tthin}))
\begin{equation}
 E_\textrm{SCU}^\textrm{SE}(t) \approx \delta(t) - \theta(t) b e^{-\tfrac{\tilde{\gamma}}{2} t} \;,
\end{equation}
with $\tilde{\gamma} = \gamma + b$.
In the opposite case of enhanced coherent excitation, a near-instantaneous displacement by half the resonant wavelength is required.
From Eqs.~(\ref{eq:ein_phase1}) and~(\ref{eq:ein_phase2}) we find
\begin{equation}
 E_\textrm{SCU}^\textrm{Boost}(t) \approx \delta(t) + \theta(t) b e^{-\tfrac{\tilde{\gamma}}{2} t} \;,
\end{equation}
such that the sign of the second pulse is flipped. 
To compute the total field in forward direction behind the target, we assume
the same target thickness $b$ for simplicity and find
\begin{eqnarray}
 E_\textrm{out}^\textrm{SE/Boost}(t) &=& \left[  \delta(t) \mp \theta(t) b e^{-\tfrac{\tilde{\gamma}}{2} t} \right] \ast \left[  \delta(t) - \theta(t) b e^{-\tfrac{\tilde{\gamma}}{2} t} \right] \nonumber \\
 &=& \delta(t)  + \theta(t) e^{-\tfrac{\tilde{\gamma}}{2}t} \left( - b \mp b \pm b^2 t \right) \;.
\end{eqnarray}
Comparing the intensities yields
\begin{equation}
 |E_\textrm{out}^\textrm{Boost}(t)|^2 - |E_\textrm{out}^\textrm{SE}(t)|^2 = 4 b^2 e^{-\tilde{\gamma} t} \left(b t -1 \right)\;.
\end{equation}
Thus, for $t< b^{-1}$ the intensity for the case of stimulated emission dominates, while for $t> b^{-1}$ the scheme with enhanced excitation results in a higher detection rate.
The same qualitative behaviour is observed in the experimental data shown in Fig.~\ref{fig:exp} and in the full theory calculations shown in Fig.~\ref{fig:exp} and Fig.~\ref{fig:patterns}.

\begin{figure}[t]
 \centering
 \includegraphics[width=0.95\columnwidth]{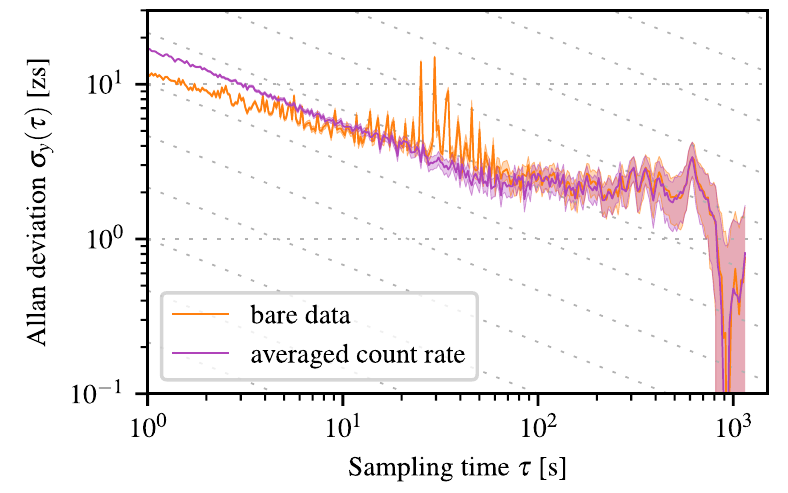}
 \caption{\textbf{Influence of detector dead time.} The orange curve shows the Allan deviation for the case of stimulated emission and a linear noise model.
 The fluctuations at intermediate sampling times are due to dead time in the data acquisition system. In the purple curve these drops in the count rate were artificially
 avoided by assuming a constant count rate instead. Consequently, the fluctuations vanish. Shaded areas are error ranges, diagonal grid lines indicate $1/\sqrt{\tau}$ scaling.}
 \label{fig:allan_deadtime}
\end{figure}

\begin{figure*}[t]
 \centering
 \includegraphics[width=1.95\columnwidth]{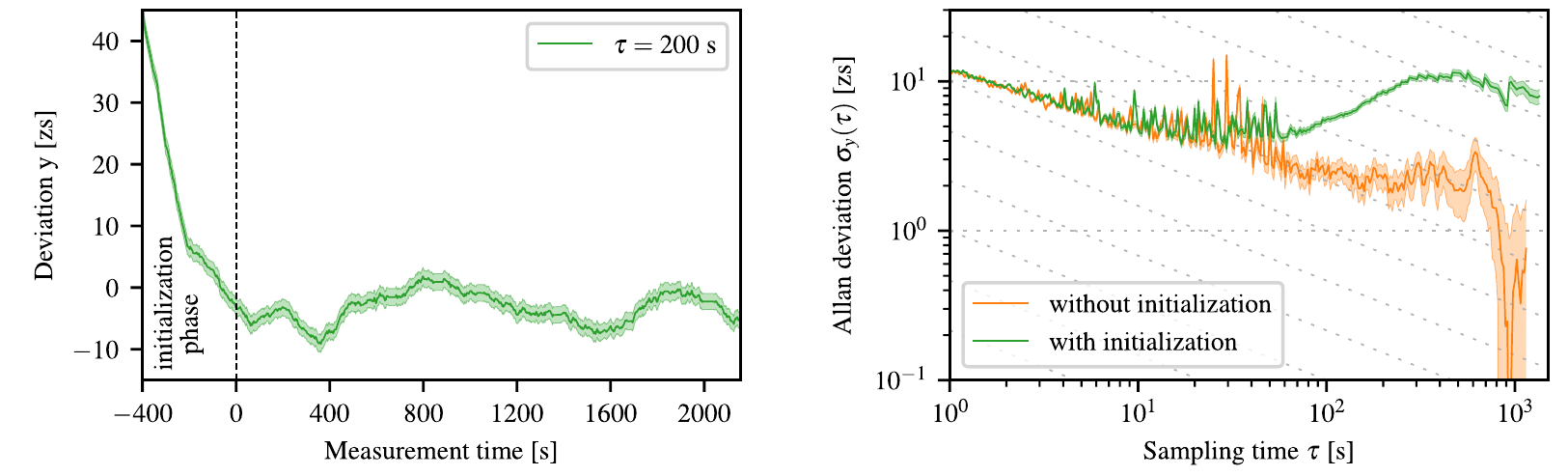}
 \caption{\textbf{Systematic deviations throughout the initialization phase.} The left panel shows the obtained temporal deviations $y_i$ for samples of sampling time $200$~s
 throughout the full measurement period for the case of enhanced coherent excitation. Note that in this panel, temporally overlapping samples were analysed in order to trace the time evolution of the deviation throughout the initial phase of the measurement. The first $\approx 400$~s during the initialization period  of the measurement time exhibit significant differences. As shown in the right panel,  inclusion of this time range  leads to systematically larger Allan deviations for sampling times above approximately $100$~s.  Shaded areas are error ranges, diagonal grid lines indicate $1/\sqrt{\tau}$ scaling.}
 \label{fig:allan_400}
\end{figure*}

\section{\label{app-multi}Multidimensional detection}
In our experiment, we make use of an event-based detection system which records, among other quantities, the time- and energy information for each photon separately. It thus provides access to a two-dimensional time- and energy-resolved dataset which contains the full holographic (amplitude and phase) information in its interference structures.
This is crucial in two respects. First, the Allan deviation analysis requires an {\it a posteriori} splitting of the data into time bins of variable duration $\tau$. This is only possible if the arrival time of each photon is stored. Second, we will show below that the time-dependent intensity, which was used in previous experiments, does not provide access to the key observables studied here, namely the complex nuclear dipole moment and the stability of the coherent control scheme.
To better appreciate the difference between our multidimensional detection and the standard time-dependent intensity measurement, it is important to note that in order to determine the nuclear dynamics, one in fact has to solve an inverse problem of extracting the  nuclear dipole moment from the scattered light. The time-dependent intensity measured in previous works does not provide sufficient information to unambiguously solve this inverse problems, which has posed a fundamental obstacle to access the matter (nuclear) part of the system.  

To illustrate the necessity of our multidimensional spectroscopy method, we consider the setup used in our experiment, with the three motions shown in Fig.~\ref{fig:multidim1}(a). Motion \mone corresponds to a rapid jump shortly after the arrival of the x-ray pulse by half the resonant wavelength $\lambda_0/2$, which would lead to the enhanced coherent excitation case. The second motion \mtwo is a similar displacement, but in the opposite direction. The third motion \mthree modifies motion \mone by an additional linear drift on top of the step-like motion. As discussed in Appendix~\ref{app-allan}, such linear drifts are the dominant source of noise expected in our setup, and the drift shown in the figure corresponds to a temporal deviation $y = 25$~zs. Our stability analysis is based on the capability to reliably detect drifts of this and smaller magnitude.  As shown in Fig.~\ref{fig:multidim1}(b,c), the three motions induce different dynamics in the target nuclei, and our experiment aims at detecting these differences. Note that somewhat counter-intuitively, motions \mone and \mtwo induce dynamics which do not only differ in phase, but also in the time-dependent magnitude of the induced dipole moments. The reason for this feature is that the two motions include opposite velocities in the approximately step-like part of the motion, leading to transient opposite Doppler shifts, and thus in turn to different spectra of the outgoing double-pulses. Thus, the target nuclei experience different driving fields. Motion \mthree differs from motion \mone by an additional drift, which translates into a corresponding additional phase dynamics of the induced dipole moments. 

Fig.~\ref{fig:multidim2}(a) show the theoretical predictions for the time-dependent intensity on resonance, which was used as an observable in previous experiments. The corresponding intensity differences obtained by subtracting the experimentally accessible intensities from each other are shown in panel (b). The results for motions \mone and \mtwo essentially coincide. Motion \mthree only  differs slightly in the depth of the beat minima and is essentially indistinguishable from the other motions, in particular if practical limitations on data acquisition are taken into account. Thus, we conclude that the time-dependent intensity alone is not capable of distinguishing key motions of relevance to our analysis from each other as a matter of principle, and therefore cannot distinguish the different nuclear dynamics induced in the target nuclei. 

The multidimensional detection technique used in our experiment provides time- and energy-resolved spectra as shown in Fig.~\ref{fig:exp}(a,b). To illustrate the advantage of this approach, we show relative intensity differences $(I_\text{\mtwo} - I_\text{\mone})/(I_\text{\mone} + I_\text{\mtwo})$ of the 2D spectra obtained for motions \mone and \mtwo in Fig.~\ref{fig:multidim2}(c). It can be seen that the two motions lead to rich systematic structure with full visibility. Therefore, through the 2D spectra one can easily distinguish the two motions, while the time-dependent intensities on resonance in panel (a) for the two motions cannot. Finally, panel (d) of  Fig.~\ref{fig:multidim2} shows intensity differences of the three motions for sections through the measured 2D spectra at particular M\"ossbauer drive detunings $\delta$. It can be seen that all three motions give rise to significant intensity differences, which furthermore exhibit characteristic time-dependencies for each detuning separately. In our data analysis, we compute two-dimensional theory spectra and compare them to the entire recorded two-dimensional spectrum at once, thereby including all M\"ossbauer detunings in a single fit.  The rich interference structures encode full tomographic (amplitude and phase) information on the light scattered by the first absorber, and lead to a strong sensitivity of the fit to the slightest deviations in the piezo motion and the nuclear dynamics.  These examples clearly show that the time-dependent intensity measured in previous experiments is incapable of distinguishing motions which are crucial to our results, in contrast to the 2D time- and energy resolved spectra recorded in our experiment.

\section{\label{app-scu}Reconstruction of the SCU motion}
The reconstruction of the SCU motion was performed based on the method in Ref.~\cite{Heeg2017}. 
In the experiment, the  duration of the periodic motional pattern $x_0(t)$ of the SCU was chosen as  a multiple of the synchrotron bunch clock period, and locked to the bunch clock. This way, stable temporal shifts between the x-ray pulses and the motional pattern could be adjusted. The target was mounted on a Doppler drive, such that the relative detuning $\delta$ between the target resonance energy and that of the nuclei in the SCU  could be tuned via the velocity $v$ of the drive. Using our event-based detection system, we recorded two-dimensional time- and velocity resolved intensities $I(t, v)$ for different temporal shifts of the motional pattern. The set of shifts was chosen in such a way that the recorded time-dependent intensities span the entire motional sequence. Each measurement covers times from $18$~ns to $170$~ns after the excitation with the initial x-ray pulse, and the velocity was recorded in the range $-0.0228$~m/s to $0.0228$~m/s. 
Using an evolutionary algorithm, we fitted the applied motional sequence to the measured data without imposing a particular model for the motion. In this step, the experimentally measured and the theoretically expected data are compared using a Bayesian log-likelihood method. For this method,  we maximized the Bayesian likelihood~\cite{bayes} under the assumption that the photon counts for each data point in $I(t, v)$ are Poisson distributed~\cite{Mandel}. For a given ideal datum $n_{\mathrm{theo},i}$ with index $i$, the probability of obtaining the experimental count number $n_{\mathrm{exp},i}$ is then 
\begin{equation}
P(n_{\mathrm{exp},i}|n_{\mathrm{theo},i}) \propto (n_{\mathrm{theo},i})^{n_{\mathrm{exp},i}} \: \frac{e^{-n_{\mathrm{theo},i}}}{n_{\mathrm{exp},i}\,!}\,. 
\end{equation}
The likelihood for the whole experimental dataset including all data points $i$ is
\begin{equation}
P(\mathrm{exp}|\mathrm{theo}) = \prod_i P(n_{\mathrm{exp},i}|n_{\mathrm{theo},i})\,. 
\end{equation}
Assuming uniform priors~\cite{bayes}, $P(\mathrm{exp}|\mathrm{theo}) \propto P(\mathrm{theo}|\mathrm{exp})$, which allows  for the determination of the most likely theoretical prediction given the experimental data. Thus, we calculate $n_{\mathrm{theo},i}$ for each motion obtained during the  evolutionary algorithm, and maximize $P(\mathrm{theo}|\mathrm{exp})$ to choose the most likely one. As a result of this  evolutionary algorithm, we obtain the full periodic motion $x_0(t)$.

\section{\label{app-allan}Stability and Allan deviation}

The stability of our control scheme is given by the stability of the relative phase between the excitation and the control pulses experienced by the target nuclei. Since the first excitation pulse interacts with the target at $t\approx 0$, this phase depends on the relative motion of SCU and target during the subsequent 176~ns of each experimental run. In contrast, drifts or perturbations in between different runs do not affect the stability. As a result of this relative dependence, in our modeling we can equivalently attribute imperfections in the stability of our setup either to noise or drifts in the relative phase, or to corresponding perturbations in the SCU motion.

To quantify the stability of our coherent control scheme, we use the Allan deviation measure~\cite{Allan1966}, which is obtained by the analysis illustrated in  Fig.~\ref{fig:allan_scheme}. The respective recorded datasets are split into non-overlapping samples with equal sampling times $\tau$. For example, for $\tau = 10$~s the first sample comprises the data taken in the time range $0-10$~s, the second sample is formed by the data recorded in the time range $10-20$~s, and so forth. For all $N$ samples obtained for a given sampling time $\tau$, we determine a quantity $y_i$ characterizing the double-pulse sequence in the interval $i$ in terms of a temporal deviation as explained below. From the $y_i$, the Allan deviation $\sigma_y(\tau)$ can be computed according to 
\begin{equation}
 \sigma_y(\tau) = \left( \frac{1}{2 (N-1)} \sum_{i=1}^{N-1} (y_{i+1} - y_i)^2 \right)^{\tfrac{1}{2}} \;. \label{eq:allan}
\end{equation}

It remains to determine $y_i$ from the experimental data as function of $\tau$. However, for short measurement intervals $\tau$, the experimental statistics is not sufficient for a full independent recovery of the applied double-pulse sequence. Therefore, we make use of the direct correspondence of the double-pulse phase and the SCU motion, and base our analysis on the SCU motion $x_0(t)$ obtained as the best fit for the entire experimental dataset. In the first step, we modify $x_0(t)$ using an error model, which depends on a model parameter specified below. In the second step, we fit the modified motion to the experimental data in each interval $i$ of duration $\tau$ separately, using the model parameter for the fit. In this fit, we use the same Bayesian log-likelihood method as for the recovery of $x_0(t)$. In the third step, we translate the best fit for the model parameter into the desired temporal deviation $y_i$ according to the error model. 

To derive an error model, we decompose the perturbation $\delta x(t)$ to the motion into frequency components as $\delta x(t) = \sum_\omega x_\omega(0) + a_\omega \,\sin(\omega t + \phi_\omega)$, taking into account offsets $x_\omega(0)$ and relative phases $\phi_\omega$ for each frequency component $\omega$ separately. For $\omega t < 1$, a series expansion yields $\delta x(t) \approx \delta x(0) + A\,t$, where $\delta x(0) = \sum_\omega x_\omega(0) + \sin(\phi_\omega)$ and $A = \sum_\omega a_\omega \omega \cos(\phi_\omega)$. Therefore, during each experimental run of 176~ns, perturbations at least for all frequencies below $\sim 2\pi/(176~\textrm{ns})\sim 10$~MHz together can be summarized into a constant offset $ \delta x(0) $ not affecting the relative phase between the two pulses, and a linear drift motion $A\,t$ randomly varying from run to run. 
Therefore, we use $x_i(t) = x_0(t) + A_i t$ as our main error model, with the free parameter $A_i$ characterizing the magnitude of the drift in each interval $i$. The parameter $A_i$ then translates into the desired temporal deviation $y_i$ as $y_i = A_i t_2 / c$, where $t_2 = 170$~ns is the maximum time of our data acquisition and $c$ is the speed of light. With this choice, the temporal deviation $y_i$ constitutes an upper bound for the phase error acquired due to the drift with parameter $A_i$.

Next to the linear drift motion, we also employed two other noise models to analyze the stability of our data. 
First, a scaling of the expected motion by a constant factor, $x(t) = (1 + s) x_0(t)$.  For example, in the case of a $\pi$ phase jump in $x_0(t)$, a scaling by $s$ corresponds to a phase shift of $s\cdot\pi$, and thus a temporal shift $s\cdot 143.5$~zs. This model, for instance, takes into account fluctuations in the voltage applied to the piezo, which to  a very good approximation  translates into a scaling of the displacement.
Second, we superimposed the base motion with a small step-like displacement, $x(t) = x_0(t) + d\cdot \theta(t-0^+)$. The phase displacement  $d$  translates into a temporal shift of $d/\pi\cdot 143.5$~zs. $0^+$ indicates a time close to zero  immediately after the excitation pulse has left the target. This model  tests for the presence of potential phase offsets between the excitation and control pulses.%

In our analysis we found that the linear model (i) constitutes the dominating type of error.
The Allan deviations for the different noise models in the case of enhanced coherent excitation are shown in  Fig.~\ref{fig:allan_noisemodels}.
While the linear noise model predicts an optimum deviation of $\sigma_y(\tau) \approx 1$~zs for the given data, the uncertainties obtained from the other two models reach well below the zeptosecond scale.

\section{\label{app-detector}Detector dead time}
In all curves shown in Fig.~\ref{fig:allan_noisemodels} as well in the curves in Fig.~\ref{fig:allan} one observes unexpected fluctuations in the Allan deviations at sampling times between $\tau \approx 10$~s and $\tau \approx 60$~s.
The cause for this is a limitation of the employed data acquisition system, which occasionally suffered from dead times of a few ten seconds, due to overload resulting from a too high signal rate. As a result, some data samples with respective sampling times contain only a few or even no counts, which spoils the determination of $y_i$ and in turn leads to large Allan deviations.
This effect can be removed in the data analysis by choosing the samples not according to equal measurement times, but according to equal counts. In other words, instead of the fluctuating count rate in the experiment with its dead times, a constant averaged count rate is assumed. As shown in Fig.~\ref{fig:allan_deadtime}, evaluating the Allan deviation with this method indeed suppresses the fluctuations at intermediate times, which shows that they originate from the detector dead time.

\section{\label{app-init}Systematic deviations throughout the initialization phase}
In the Allan deviation shown in Fig.~\ref{fig:allan}, it is not fully clear if the experimentally achieved stability has already reached its limit, and only an upper bound for possible systematic effects can be given.
In order to interpret this result and to verify our analysis, we artificially introduced systematic  deviations, by recording spectra already during the initial time after starting the piezo motion, before the piezo reached stable thermal and mechanical conditions. In this initial time, systematic drifts in the deviation $y$ as a function of the measurement time may occur. The corresponding results for samples with sampling time $200$~s are shown in the left panel of Fig.~\ref{fig:allan_400}  over the full measurement period, including the initialization phase. Note that in this plot, temporally overlapping samples were analysed, in order to trace the time evolution of the deviation with a high temporal resolution. For example, the first deviation is calculated from data in the time range $0-200$~s, the next deviation for the range $1-201$~s, and so forth. We find that $y$ systematically drifts for an initial period of about 400~s. Afterwards, only small residual fluctuations are observed over the remaining measurement time. In the right panel of Fig.~\ref{fig:allan_400}, the Allan deviations with and without this initial phase are compared. It can be seen that the initialization leads to a clear systematic trend of the Allan deviation as compared to the case without the initial phase: the Allan
deviation begins to increase again for sampling times exceeding approximately 100~s, which is the expected behaviour in case of systematic drifts.

\section{\label{app-samples}Samples}
As resonant nuclear sample we used a single-line stainless-steel foil (Fe\textsubscript{55}Cr\textsubscript{25}Ni\textsubscript{20}), with iron enriched to about 95\% in \textsuperscript{57}Fe and
with thickness 1~{\textmu}m.
The x-ray double-pulse sequence was created using an \textalpha-iron foil with thickness of $\approx$~2~{\textmu}m,
also enriched in \textsuperscript{57}Fe. An external magnet was used to align its magnetization and the
setup was arranged such that only the two $\Delta m = 0$ hyperfine transitions of the 14.4 keV
resonance in \textsuperscript{57}Fe were driven.
To displace the \textalpha-iron foil we employed a piezoelectric transducer consisting of a polyvinylidene
fluoride (PVDF) film (thickness 28~{\textmu}m, model DT1-028K, Measurement Specialties, Inc.).
The piezo was glued on a plexiglas backing and was driven by an arbitrary function generator (model Keysight
81160A-002).

\bibliographystyle{naturemag}
\bibliography{pumpprobe}

\end{document}